\RequirePackage{fix-cm}
\documentclass[smallextended]{svjour3}       
\smartqed  
\usepackage{graphicx}
\usepackage{dcolumn}
\usepackage{bm}
\usepackage{amssymb}
\usepackage{amsmath}
\usepackage{amsfonts,mathrsfs}

\begin{document}

\title{Forewords for the special issue `Pilot-wave and beyond: Louis de Broglie and David Bohm’s quest for a quantum ontology'}


\author{Aur\'elien Drezet}


\institute{A. Drezet \at Institut NEEL, CNRS and Universit\'e Grenoble Alpes, F-38000 Grenoble, France \\
            \email{adrezet@neel.cnrs.fr} }

\date{Received: date / Accepted: date}

\maketitle

\begin{abstract}
In order to celebrate this double birthday the journal Foundations of Physics publishes a topical collection “Pilot-wave and beyond” on the developments that have followed the pioneering works of Louis de Broglie and David Bohm on quantum foundations. This topical collection includes contributions from physicists and philosophers debating around the world about the scientific legacy of Bohm and de Broglie concerning the interpretation and understanding of quantum mechanics.
In these forewords we give a general review of the historical context explaining how de Broglie and Bohm developed their interpretations of quantum  mechanics. We further analyze the relationship between these two great thinkers and emphasize the role of several collaborators and continuators of their ontological approach to physics. 

\end{abstract}

\section{Introduction}\label{sec1}
\indent  
Seventy years ago the  main-stream and leading journal \emph{Physical Review} published two articles by David Bohm with the titles `A suggested interpretation of the quantum theory in terms of hidden variables (parts I and II)'~\cite{Bohm1952a,Bohm1952b} submitted together the 5$^{th}$ of July 1951 and published the 15$^{th}$ of January 1952 in the same volume. It is true to say that these two articles had, on the long term, a tremendous effect on the scientific community. Without this work by Bohm John Bell would probably not have even considered the nonlocality issue as a serious problem \footnote{In a article published in Foundations of physics for celebrating Louis de Broglie ninetieth birthday Bell wrote: `In  1952 I saw the impossible done. It was in papers by David Bohm'\cite{Bell}. }~\cite{Bell1} and as a consequence he would not have discovered his famous theorem  `On the Einstein Podolsky Rosen Paradox'~\cite{Bell2}.  Subsequently, if after Bohm, Bell would not have written his corners stone articles we can speculate that neither  John Clauser,   Alain Aspect, nor Anton Zeilinger would  have started their experimental projects; and therefore would not have won the Nobel prize this year `for experiments with entangled photons, establishing the violation of Bell inequalities and pioneering quantum information science'.\\
\indent   Nevertheless,  and despite all the scientific achievements and technological implications based  on quantum entanglement it is also fair to say   that if everything started with some fundamental questions concerning reality and hidden variables   the work of Bohm was badly received in the 1950's at a time when even asking if hidden variables could exist and explain the underlying quantum reality was considered as an anathema. For the community David Joseph Bohm became immediately one of those heretics, like Einstein or Schr\"odinger, criticizing the orthodox Copenhagen quantum interpretation.  This story is well known and documented by historians and doesn't  need to be commented too much here~\cite{Peat,Freire}.\\
\indent More interesting for us is to understand the evolution and transformation of Bohm's ideas during the 1950's. In 1951, while he was still an assistant professor at Princeton university, Bohm published a remarkable textbook on quantum theory~\cite{Bohm1951}. Compared to many other textbooks of the same technical level the version of Bohm   contains lenghtly philosophical discussions concerning materialism, determinism, causality and probability. The book is completely orthodox in its content and attempted to justify the inevitability of Bohr's interpretation and the impossibility of preserving the classical dream of a self-consistent and realist description of matter moving in space-time. Importantly, the book contains a `proof that quantum theory is inconsistent with hidden variables' obtained in relation with  the Einstein Podolsky Rosen (EPR) paradox \cite{EPR} and the principle of complementarity of Bohr. Bohm therefore concluded on the nonexistence of `hidden variables underlying quantum mechanics'. Albert Einstein was very much interested by the book and gave a phone call to Bohm to discuss with him. The content of this discussion was summarized by Bohm himself in an interview made by his friend Maurice Wilkins in 1980 \cite{BohmWilkins1980}, the story is also recounted in Bohm's  biography written by David Peat \cite{Peat} (see also \cite{PeatBohm} and \cite{Bohm1982} which include other recollections of the whole story by Bohm, and finally Max Jammer book and article \cite{Jammer,Jammer2}). After reading the book Einstein discussed with Bohm. We don't exactly know what was the precise content of these discussions  (very probably it concerned the EPR paradox \cite{EPR}) but afterwards   Bohm decided  to change his mind~\cite{Peat,Freire}. Soon, he developed an alternative realist interpretation of quantum mechanics and wrote a first shorter draft of the manuscript published in 1952~\cite{Bohm1952a,Bohm1952b} in two parts. Einstein and Pauli read this manuscript and reacted promptly.  Pauli~\cite{BohmPauli1951} was particularly critical and dismissive and pointed out that this work  has been already proposed by Louis de Broglie in 1927 under the name `pilot-wave theory' and  was discussed in details during the $5^{th}$ Solvay conference in Brussels  where Pauli, Einstein  and others already emphasized key difficulties with the ideas of de Broglie~\cite{Brussels,Valentini}. Einstein also stressed the priority of de Broglie and as a consequence  Bohm contacted de Broglie in Paris.\\
\indent It is actually remarkable that Bohm didn't know about de Broglie early work.  De Broglie  published an article in 1927 \cite{debroglie1927} and a book in 1930 \cite{debroglie1930} that detailed the pilot-wave interpretation presented in 1927 and both works have been translated in english immediately. Moreover, in 1928  Kennard also published an article in the Physical Review about a similar interpretation~\cite{Kennard1928} in term of particle trajectories in the configuration space. While the article of Kennard doesn't mention the work of de Broglie (but refers to the hydrodynamical work by Madelung~\cite{Madelung1926}) the article contains a detailed discussion of some measurement  protocols in quantum mechanics within this formulation (this also constitutes a fundamental contribution of the Bohm's articles). Even more remarkable, Nathan Rosen, the close collaborator of Einstein, co-author of the EPR paper, published  in 1945 an article~\cite{Rosen1945} reproducing many elements of the pilot-wave theory for a single particle and this without even quoting de Broglie (Rosen paper gives a reference to the work of Madelung \cite{Madelung1926} and Kennard \cite{Kennard1928}). Because of his connection with Einstein, it is apriori not impossible that Bohm could know about these works of Rosen and Kennard (a textbook by Kennard on quantum mechanics is quoted in \cite{Bohm1951}).\\
\indent While  we will probably never know the details of the story   it is important to mention that de Broglie actually abandoned his interpretation already in 1930 as he explained in his book~\cite{debroglie1930}.  De Broglie provided several reasons for this. First, the  particle dynamics  proposed by de Broglie leads to curious and counterintuitive features such as particle not moving in a central potential. Pauli objected that  the theory cannot deal well with problems involving several particles in  a given region of space. De Broglie (and also Leon Brillouin) actually responded correctly to the objections but the negative reception was very demotivating for de Broglie. Furthermore, and this is  probably more important, de Broglie felt that the physical meaning  of the guiding wave $\Psi(t,\mathbf{x}(t))$ guiding the particle with trajectory $\mathbf{x}(t)$  and de Broglie-Bohm velocity 
\begin{eqnarray}
\frac{d}{dt}\mathbf{x}(t)=\frac{\hbar}{m}\textrm{Im}[\frac{\boldsymbol{\nabla}\Psi(t,\mathbf{x}(t))}{\Psi(t,\mathbf{x}(t))}]
\end{eqnarray}
was unclear. Indeed, in the orthodox interpretation   the wave is told  to collapse during a measurement and therefore the wave cannot have any physical action on any other system afterwards (this was already the point made by Einstein at Solvay's conference~\cite{Brussels,Valentini}).  This nonlocal feature is very mysterious and  de Broglie could not accept this conclusion.  If the pilot-wave theory agrees with the results obtained with the usual interpretation proposed by   Bohr, Born, and Heisenberg   it must be able to physically explain this collapse of the wave function. Indeed, such a collapse is not just an updating of information unlike in classical probability theory: The reason being that waves can interfere and therefore the $\Psi$-wave must also have an ontological (ontic) nature.  This actually implies to clarify  the meaning of `empty waves' and nonlocal interactions in the pilot-wave theory and de Broglie felt this was impossible at that time. A different reason for de Broglie abandonment was related to the probabilistic interpretation developed by Max Born where any return to determinism was considered as a regression. The subsequent discovery of the uncertainty principle by Heisenberg confirmed this tendency: the classical determinism \`a la Laplace was definitively outmoded (Born's ideas were also clearly motivated by the fact that strong determinism \`a la Laplace apparently contradicts the existence of some form of `free-will' at the atomic level \footnote{Issues concerning acausality, free-will or even rationalism in the works of Born, Heisenberg, Jordan or Pauli have been discussed by historians and sociologists of sciences (see for example \cite{Beller,Forman}) and by physicists \cite{Selleribook,Bricmont,Becker}. In the particular case of Max Born, it must be emphasized that in agreement with Cassirer he distinguished between a subjective free-will applied to human affairs and a form of atomic free-will or freedom associated with acausality at a more fundamental quantum level \cite{Bornbook}. Interestingly, in his first work of 1926 about the statistical interpretation of quantum mechanics Born let as a philosophical question the fundamental nature of undeterminism~\cite{Bornpaper}. Moreover, in a work published in French and German in 1958-59 he defended the view that classical physics was not completely deterministic (in order to dismiss the theories proposed by Bohm and Vigier) and emphasized that he had always disliked strong determinism \`a la Laplace that, obviously, constrasts with the `uncertainties that dominates human life and thought'~\cite{Bornarticle} (for a similar discussion see also \cite{Selleribook}).   }).  The dream of de Broglie of proposing a deterministic theory reproducing exactly the probabilistic predictions of quantum mechanics looked impossible since for many philosophers and scientists the trend of the century was away from determinism and a step backward was looking unlikely.  Moreover, there is a final reason for de Broglie abandonment of the pilot-wave approach:  Indeed, already in 1925 de Broglie attempted to develop a more complicated theory in which the wave would be unified with the particle in the 4D space time and not in the configuration space.  In this approach known as the double solution theory~\cite{debroglie1925a,debroglie1925a,debroglie1926,debroglie1926a,debroglie1926b,debroglie1927,debroglie1928}  the particle is actually a kind of localized defect or singularity surfing atop a physical base wave $u(t,\mathbf{x})$.  Inspired by older ideas made by Einstein for the photon the singularity is in his theory guided  by the base wave (like the point-like particle by the Schr\"odinger wave in  the pilot-wave approach) but the two objects are now non-linearly coupled and synchronized and constitute what we nowadays call a soliton or solitary wave propagating as a whole. The double solution theory, which is strongly related to Einstein's quest for a unified (classical) field theory, motivated  the pilot-wave model but its mathematical complexity blocked de Broglie and he subsequently abandoned this project in 1928.\\  
\indent This was so until he received the letter of Bohm in 1951. De Broglie reacted promptly by showing his priority and explained to Bohm that the pilot-wave theory was problematic and must be abandonned. He even published a short paper in 1951 before the publication of Bohm paper criticizing the whole pilot-wave strategy~\cite{debroglie1951}. This created quarrels between the two men (the story is well documented in the recent work of Besson \cite{Besson}). The tension decreased when de Broglie changed his mind and under the influence of his young assistant Jean Pierre Vigier decided to restart the double solution research program. In the following  decade de Broglie  and Vigier developed a research group on the the double solution theory and Bohm and Vigier collaborated on several issues concerning the pilot-wave theory. In particular, the theory was subsequently extended to particles with spin $1/2$ and 1 involving the relativistic Dirac and Duffin-Kemmer-Petiau equations~\cite{debroglie1952,Vigier1952,deBroglie1956,Vigier1956}. One of the central question that was debated by Bohm and Vigier concerned probability and the Born rule. How indeed can we justify the probability formula $|\Psi|^2$ of quantum mechanics from an underlying deterministic theory? In 1926 de Broglie showed that if the fluid density obtained from the wave equation can be identified at one time with a probability density for the presence of particle then this will true at any other time in agreement with Liouville theorem~\cite{debroglie1926}. Moreover, Pauli and other criticized this point and as a consequence  Bohm and Vigier attempted to derive a mechanism for forcing the Born rule to be a kind of statistical attractor during complex interaction processes. They speculated on the existence of a sub-quantum thermal bath interacting with the particles and generating a Brownian motion superposed to the stream flow given by the deterministic de Broglie-Bohm pilot-wave dynamics \cite{Bohm1953,Bohm1954,Vigier1956}. It is important therefore to point out that neither Bohm  nor Vigier were strict adepts of determinism. Better, they emphasized the preeminence of causality. Strongly influenced by Marxist ideas going back to Lenin  they  both stressed the role of an `infinite number of levels' for describing the material world~\cite{Bohm1957,Vigier1961}. Therefore, they pictured a kind of fractal universe where at each description level a stochastic-fluctuating dynamics (superposed to a deterministic mean motion law) is needed to characterize effectively the underlying  and hidden subquantum levels. These ideas were actually defended by Bohm already in 1951, as confirmed by the recent discovery of a old manuscript  written by Bohm in 1951 and sent to de Broglie the same year which has the title `A causal and continuous interpretation of the quantum theory' \cite{Drezet2021}. It is also interesting to point out that the new conceptions of Bohm were not completely at variance with his own older ideas presented in his book of 1951. Indeed, in~\cite{Bohm1951} p.~29 Bohm already wrote that there is a unlikelihood of completely deterministic laws on  a deeper level. While the reason for this unlikelihood proposed in his 1951 textbook was related with the uncertainty principle and the EPR paradox the new   `infinite number of levels' approach based on an Einsteinian realistic framework always conserved a strong contact with the complementarity principle of Bohr. Bohm analyzing Bohr's conceptions explained why it was natural for him to `interpret the indeterministic features of the quantum theory as representing irreducible lawlessness; for, because of the indivisibility of the experimental arrangement as a whole, there is no room in the conceptual scheme for an ascription of causal factors which is more precise and detailed than permitted by the Heisenberg relations'~\cite{Bohm1980} p.~97. In the same book he wrote `the view of the world as being analogous to a huge machine, the predominant view from the sixteenth to nineteenth centuries, is now  shown  to be only approximately correct'~\cite{Bohm1980} p.~167. However, in his quest for a realistic ontological description of quantum theory  Bohm found an alternative to Bohr extreme and pessimistic diagnostic still preserving in some sense the quantum wholeness and indivisibility of the quantum world. In a late interview for the BBC radio, Bohm explained that  `We should say that quantum mechanics doesn't explain anything; it merely gives a formula for certain results. And I'm trying to give an explanation'~\cite{Ghost} p.~127.\\ 
\indent In a sense  Bohm and also de Broglie  wanted to go beyond the usual resignation associated with the Copenhagen interpretation. Instead of accepting Bohr's dictum that one cannot understand the quantum world using a clear causal picture in space and time they were looking for a more classical mode of understanding where we could not say:  Look guys there is a {\it Terra Incognita} on our physical map of the quantum Universe, but unfortunately we cannot go beyond that point. The aim for them was thus to go beyond the rethoric of inevitability associated with antirealism and acausality; in other words, to find a substitute to `wave-particle duality'  and to replace it by a clear coexistence of wave and particle associated with a specific dynamical law. In this context the analysis of quantum measurements is particularly relevant and crucial for Bohm. In his 1952 article Bohm wrote: `We differ from Bohr, however, in that we have proposed a method by which the role of the  apparatus can be analyzed and described in principle in a precise way, whereas Bohr asserts that a precise conception of the measurement process is as a matter of principle unattainable'~\cite{Bohm1952b}. \\
\indent Moreover, over the years some differences of analysis and appreciation between de Broglie, Bohm, and Vigier weakened the collaboration.  De Broglie, and also Vigier, wanted to preserve the goal of a mechanistic description of the world involving a hydrodynamical description of a fluid with subquantum fluctuations (de Broglie developed the so-called thermodynamics of the isolated particle~\cite{debrogliethermo} and Vigier developed several stochastic models inspired partly by Nelson theory~\cite{VigierJeffers}) whereas Bohm felt that the quantum potential $Q=-\frac{\hbar^2}{2m}\frac{\Delta |\Psi|}{|\Psi|}$ which depends  essentially on the form of the wave function  $\Psi$ represents a non-mechanist description of nature associated with an `active information'~\cite{BohmHiley}. In the end Bohm was unconfortable with the double solution of de Broglie.  Bohm concluded that contrarily to his own approach involving the quantum potential and active information `it will not be possible to obtain  solutions of the [double solution] field equations which would lead to the very great accelerations that are in general implied by the guidance relation'~\cite{BohmHiley} p.~39.  Moreover, the biggest issues concerned perhaps the importance of nonlocality.  De Broglie, following Einstein, disliked very much the concept in conflict with the spirit of relativity and tried to develop a version of the double solution that was fundamentally local. With few collaborators and students (Fer, Lochak, Andrade e Silva~\cite{Silva,debroglieSilva} etc..) he continued to work on his quest of a local version of the double solution theory but without never finding a satisfactory model. For Vigier and Bohm the difficulties with the double solution lay in the nonlocality. They both accepted and welcomed the consequences of Bell's theorem and faced the challenge of  understanding its implication either by using stochastic models for a `Dirac Ether' involving instantaneous connections~\cite{VigierJeffers}, or by developing further the physics and metaphysics associated with the quantum potential for quantum fields and Dirac particles in the relativistic domain~\cite{BohmHiley}.\\
\indent  After the time of the pioneers and quantum dissidents a re-surge of interest in the pilot-wave theory started in the late 1970's when two students Chris Philippidis and Chris Dewdney, under the supervision of Basil Hiley, developed the first numerical calculations of de Broglie-Bohm trajectories in iconic examples such as the double-slit experiment and the tunnel effect \cite{Chris1979,Chris1982}. The visual impact of these results on Bohm himself and on a larger public was clear, and Bohm and Vigier groups published several papers on the understanding of  particles trajectories in paradigmatic experiments such as Wheeler delayed choice~\cite{Chris1985},  EPR/Bell scenario~\cite{Chris1988}, or Stern-Gerlach measurements~\cite{Chris1986}.  The publication in 1993 of the remarkable textbooks: `The quantum theory of motion' by Peter Holland \cite{Holland} and, the `Undivided Universe' by Hiley and Bohm~\cite{BohmHiley} containing many discussions, calculations, and numerical illustrations also played an important role in the dissemination and popularization of the theory.\\
\indent Nowadays, several groups of physicists around the world work with the de Broglie-Bohm theory using different motivations, strategies and methods. The differences in methods and strategies are related to the questions already asked by the pioneers de Broglie and Bohm and their collaborators concerning probability, locality, causality, or extensions of the theory to the relativistic domain (e.g. involving quantum fields). This naturally challenges the claim often found in the literature that  `Bohmian mechanics', i.e., the de Broglie-Bohm pilot-wave theory is just a re-interpretation of quantum mechanics rigorously empirically equivalent to  the standard  `interpretation'. Clearly, it is indeed possible to build formally a version of the pilot-wave theory reproducing exactly quantum mechanics (this is often named `Bohmian mechanics'). However, important questions  concerning the role of empty waves or the Born rule (i.e., related to the justification of the `quantum equilibrium regime') show that some models based on de Broglie and  Bohm approaches can predict new results (see for example the book by Selleri~\cite{Selleribook} where some controversial  but interesting issues concerning empty-waves are defended).  Some of these possibilities were already discussed by de Broglie, Bohm and their collaborators and should not be dismissed too easily. Perhaps such approaches would play a role in researches for unifying   quantum mechanics and general relativity? This was for example the hope of Vigier and de Broglie concerning the double solution program.  In this context, several historians or philosophers of science are now analyzing and questioning the meaning of the wave function  and the importance of determinism in relation with the pilot-wave approach (see for examples \cite{Cushing,Albert,Maudlin} and also \cite{Bricmont,Becker,Smolin,Valentini}). Therefore, the importance and diversity of the various topics considered in the literature clearly motivate the present special issue of the journal Foundations of Physics in order to celebrate the publication of the original Bohm's articles of 1952. This is even more justified  remembering that de Broglie and Bohm were past board members of the journal from the beginning of its creation, and contributed in several occasions with important publications (see for examples \cite{FOP1,FOP2}).\\
\indent As you will see reading this special issue, the contributors are touching all the important aspects of the de Broglie and Bohm work concerning the foundations of quantum mechanics based on the pilot-wave or double solution.  The special issue contains articles by known specialists in the fields having deeply contributed in the past and in the recent years to the understanding, development, and generalization of de Broglie Bohm theories. It also includes valuable contributions by physicists and philosophers assessing some aspects of de Broglie Bohm heritage, or trying to extend it to other ontological frameworks. Altogether, this issue of Foundations of Physics untitled {\it Pilot-wave and beyond: Louis de Broglie and David Bohm’s quest for a quantum ontology} offers to the readers a broad perspective to appreciate the importance of de Broglie and Bohm heritage in the $20^{th}$ and $21^{th}$ centuries.

\section{Acknowledgments} 
I wish to thank all the participants and contributors of this special issue. That was a great project to gather so many different perspectives concerning the impact and meaning of de Broglie Bohm theories.   I also wish to thank Carlo Rovelli and Fedde Benedictus for continuously supporting this project. As a personal note, I would like to dedicate this issue to the memory of Detlef D\"urr who was one very strong advocate of the de Broglie Bohm ontologies~\cite{Durr}, and also to the memory of Georges Lochak one of the last great collaborator of Louis de Broglie~\cite{Lochak}.

\end{document}